# Proposed cavity Josephson plasmonics with complex-oxide heterostructures

Y. Laplace[1,2], S. Fernandez-Pena[3], S. Gariglio[3], J.M. Triscone[3], A. Cavalleri[1,2,4]

[1] Max Planck Institute for the Structure and Dynamics of Matter, Hamburg, Germany

[2] Center for Free Electron Laser Science, Hamburg, Germany

[3] Département de Physique de la Matière Condensée, University of Geneva, 24 Quai Ernest-Ansermet, 1211 Genève 4, Switzerland

[4] Department of Physics, Oxford University, Clarendon Laboratory, Oxford, United Kingdom

## Abstract

We discuss how complex-oxide heterostructures that include high-$T_c$ superconducting cuprates can be used to realize an array of sub-millimeter cavities that support Josephson plasmon polaritons. These cavities have several attractive features for new types of light matter interaction studies and we show that they promote "ultrastrong" coupling between THz frequency radiation and Josephson plasmons. Cavity electrodynamics of Josephson plasmons allows to manipulate the superconducting order-parameter phase coherence. As an example, we discuss how it could be used to cool superconducting phase fluctuations with light.

Light in optical cavities has been extensively used to dress quantum states of matter, opening up fields as diverse as optomechanics, exciton-polariton condensation and Cavity Quantum Electrodynamics at the single atom-photon level. However, this extraordinary degree of control has not been applicable to many complex condensed matter systems of current interest, such as High $T_c$ superconducting cuprates. These materials are in fact often grown in bulk crystals difficult to process and to incorporate in photonic devices.

Recently, complex oxide heterostructuring has shown to be a formidable platform for engineering new electronic states of matter, as exemplified by the emergent phenomena appearing at oxide interfaces[i]. These advances in materials synthesis, based either on Pulsed Laser Deposition or on Molecular Beam Epitaxy, have improved the quality of materials to a level comparable to semiconductor heterostructures. In this paper we address how these new capabilities can be applied to a new class of cavity electrodynamics experiments that control unconventional quantum orders. As an example, we propose the design of a cavity aimed at optically dressing, manipulating and even cooling the order parameter phase of a high-$T_c$ superconductor. This could be important because cuprate superconductors are strongly affected by the loss of phase coherence near the transition temperature $T_c$, with a finite Cooper pair density thought to persist up to temperature scales in excess of $T_c$ [ii] [iii] [iv]. Hence, an important challenge involves the suppression of such phase fluctuations.

High-$T_c$ cuprate superconductors are layered materials, with superconducting planes coupled by Cooper pair tunneling across non-superconducting regions. The excitation of the relative phase between the layers ($\varphi_n = \chi_{n+1} - \chi_n$ , $\varphi_{n-1} = \chi_n - \chi_{n-1}$, etc.. see figure 1a) corresponds to collective plasma oscillations of the superconducting condensate. These modes are known as Josephson plasma oscillations, or Josephson plasmons [v][vi]. Josephson plasmons have a characteristic resonance frequency $\nu_p$, the Josephson Plasma resonance (JPR), in the GHz to THz frequency range depending on the cuprate family.

In the present case, the control of these low frequency modes is not to be realized with light fields at optical frequencies [vii,viii], which would destroy the condensate by exciting quasiparticles across the superconducting gap 2Δ. Rather, radiation at THz frequencies will be used to affect the strength and spatial distribution of the condensate phase without changing the density of Cooper pairs[ix][x][xi][xii][xiii]. As the cavity mode must couple to the interlayer Josephson plasma oscillations, the radiation should propagate with the electric field polarized perpendicularly to the superconducting planes. Also, in order to apply cavity cooling techniques such as sideband cooling [viii], the free spectral range of the cavity must be larger than the typical fluctuation frequency, that is in the THz-frequency range, hence the cavity should have sub-millimeter dimensions. The cavity design based on oxide heterostructures and shown schematically in figure 1 would fulfill all these requirements.

Let us consider a complex oxide heterostructure involving a high-$T_c$ superconducting thin film (e.g. $La_{2-x}Sr_xCuO_4$), cladded between lattice-matched insulating films ($La_2CuO_4$), with a conducting oxide at the bottom (e.g. $SrRuO_3$) and capped with a metal patch (e.g. gold) (figure 1b). This design can be realized with standard c-axis grown thin film heterostructuring, in combination with etching techniques[i].

Due to the strong sub-wavelength confinement of the electromagnetic field between the top and bottom metals (D≪ $\lambda$), resonant modes that propagate along the film can be excited. These are highly localized within the volume defined by this mesa, and have their electric field polarized along the z(c)-direction[xiv] (see figure 1c). This heterostructure corresponds then to a periodically structured array of cavities with sub-millimeter spacing, which greatly enhances the coupling efficiency of the incoming radiation as compared to a single cavity [xv].

Let us first consider a "bare" cavity without the superconducting film and based on $Au/La_2CuO_4/SrRuO_3$. The reflection spectrum of a periodic array of such cavities was computed here using the Fourier modal method [xvi]. We consider radiation impinging at normal incidence ($k_x=k_y=0$, $k_z=2\pi\nu/c$) with the electric field polarized along the y-direction (TM polarization). We modelled the optical properties of the different materials from experimental results found in the literature, including anisotropy and dissipation ($La_{2-x}Sr_xCuO_4$ [xvii] [xviii] [xix] [xx] [xxi], $SrRuO_3$[xxii] and Au[xxiii]). The computations are performed at fixed thickness D=1.2µm between the Au patch and $SrRuO_3$ and fixed periodicity p=33µm. The thicknesses of the Au patch and $SrRuO_3$ are 200nm and 500 nm respectively. The results for these bare cavities are shown in figure 2a.

For a given patch width $w$, the reflectivity as a function of frequency displays a lorentzian dip centered at the cavity resonance frequency $\nu_c(w)$ (fig 2a left panel). Because of the strong impedance mismatch at the patch edges, each patch acts as a cavity, with resonance frequency $\nu_{n,m} = c/(2\sqrt{\epsilon_\infty}w) \sqrt{n^2 + m^2}$ ($n$ and $m$ are integers denoting the number of nodes of the electric field of the cavity mode in the x and y direction respectively, $\epsilon_\infty \approx 27$ is the dielectric constant of the insulating $La_2CuO_4$[xvii] and $w$ is the width of the top metallic square patch) [xiv]. Hence, this resonance corresponds to the $(n,m) = (0,1)$ fundamental mode of the cavity, whose electric field distribution is shown on figure 1c. In fig 2a, we show the reflectivity of the heterostructure as a function of frequency and $w$ (note that on the abscissa, the patch width $w$ has been converted to the "inverse patch width" frequency defined by the relation $\nu(w) = c/(2\sqrt{\epsilon_\infty}w)$). We see that the cavity resonance frequency scales linearly with the inverse patch width, as expected from the formula above for $\nu_{n,m}$.

In figure 2b, we show a similar calculation performed on a cavity that includes a film of the optimally doped high-$T_c$ superconductor $La_{2-x}Sr_xCuO_4$ (x=17%, $T_c$=36K, $\nu_p$ = 2.3THz). The anisotropic dielectric permittivity of $La_{2-x}Sr_xCuO_4$ is expressed as a diagonal dielectric tensor. Although the model realistically includes anisotropies and dissipation, the results of the computation are dominated by the electrodynamics of the c-axis superconducting component, whose dielectric permittivity in this frequency range can be described by $\epsilon_c^{SC} = \epsilon_\infty \left(1 - \frac{\nu_p^2}{\nu^2}\right)$, with $\epsilon_\infty \approx 27$ and $\nu_p = 2.3$THz. As shown in fig2b, by tuning the fundamental mode of the cavity $\nu_c(w)$ across the JPR frequency $\nu_p$=2.3THz we observe a clear avoided crossing of the two bare dispersions $\nu = \nu_p$ and $\nu = \nu_c(w)$, which indicates strong coupling between the cavity field and the JPR.

At the frequencies for which the JPR and the cavity are resonant ($\nu_c = \nu_p$), the reflectivity displays two absorptions of similar intensities whose frequencies are separated by an amount $2\Omega$. This frequency splitting can be thought of as a Rabi splitting [xxiv] and quantifies the rate at which energy is exchanged between the electromagnetic field and the Josephson plasmon.

The behavior of such heterostructures can be quite generally captured by a simple effective medium model [xxv]. In our case, the effective dielectric permittivity $\epsilon_{eff}(\omega)$ of the insulator/high-T$_c$/insulator heterostructure can be written as $\frac{1}{\epsilon_{eff}(\omega)} = \frac{1-f}{\epsilon_\infty} + \frac{f}{\epsilon_c^{SC}(\omega)}$ ($f$ is the filling fraction, defined as the ratio between the thickness $d$ of the superconducting film and the total thickness $D$ of the cavity : $f = \frac{d}{D}$ ($0 \leq f \leq 1$)). For a cavity tuned at $\nu_c(w)$, the frequency of the hybridized resonances are given by $\epsilon_{eff}(\nu)\nu^2 = \epsilon_\infty \nu_c(w)^2$, leading to the two branches $\{\nu_+(w), \nu_-(w)\}$ plotted on fig 2b for the corresponding filling fraction $f \approx 17\%$ (black solid lines). We see that the effective medium model matches well the results of the full numerical computation.

We next analyze the strength of the coupling, evidenced by the size of the Rabi splitting. In figure 3a, we compute the reflectivity as a function of filling fraction $f$ for a cavity tuned at $\nu_c = \nu_p$. Already for relatively small filling fractions (as small as ~3%), the Rabi splitting $2\Omega$ of the two hybridized modes is observable, and exceeds the linewidth broadening originating from the cavity dissipation and JPR damping. In fig3b, we show the frequency splitting $2\Omega$ normalized to the bare cavity frequency $\nu_c = \nu_p$ yielding the normalized Rabi splitting. We observe that, even at small filling fractions, coupling in this heterostructure can be characterized as "ultrastrong" according to the usual prescription of a normalized Rabi splitting greater than ~20%.

Note that the optical properties of a dissipationless and isotropic optical excitation can be described as: $\epsilon = \epsilon_\infty(1 - \nu_p^2/(\nu^2 - \nu_0^2))$, where $\nu_0$ is the frequency of the transition and $\nu_p$ is its oscillator strength. When a material carrying this optical excitation is placed inside the cavity with a filling fraction $f$, a Rabi splitting of $2\Omega$ is observed when the cavity is tuned at $\nu_c = \sqrt{\nu_0^2 + \nu_p^2}$, yielding a normalized Rabi splitting given by

$$\frac{2\Omega}{\nu_c} = \sqrt{1 + \sqrt{f}\frac{\nu_p}{\sqrt{\nu_p^2 + \nu_0^2}}} - \sqrt{1 - \sqrt{f}\frac{\nu_p}{\sqrt{\nu_p^2 + \nu_0^2}}}$$

obtained from the effective medium model described previously.

Light-matter interaction studies have to date been realized almost exclusively for cavities tuned to optical frequencies and coupling to low oscillator strength excitations, that is for $\nu_0 \gg \nu_p$, such as the case of interband transitions in quantum wells [xxiv]. Hence, typical normalized Rabi splittings of $\sim\sqrt{f}\frac{\nu_p}{\nu_0} \ll 1$ fall in the weak to moderate coupling regime. Recently, high-density 2-dimensional electron gases with electronic transitions of $\nu_0$ and $\nu_p$ both in the THz frequency range have reached the ultra-strong coupling regime [xxv][xxvi].

Here, for the superconducting material, $\nu_0 = 0$, as the condensate response peaks at zero frequency and is characterized by the zero-frequency delta function in the conductivity. Hence, the normalized Rabi splitting reduces to $\sqrt{1 + \sqrt{f}} - \sqrt{1 - \sqrt{f}}$, and unlike the more conventional cases is independent of either $\nu_0$ or $\nu_p$. For realistic filling fractions, the ultra-strong coupling regime is easily reached.[xxvii]

So far, we analyzed the linear electrodynamical properties of these oxide heterostructures and described how the cavities can be used to optically dress the Josephson plasmons. In the following, we discuss how such cavities can as well cool thermally excited Josephson plasmons in high-Tc superconductors with cavity cooling techniques, as done in the case of a single Josephson junction [xxviii].

The electrodynamics of this cavity/high-T$_c$ heterostructure is intrinsically nonlinear due to the Josephson relation between the current density and the phase difference between the planes: $j_n = j_c\sin(\varphi_n) = j_c \sin\left(\frac{2ed^*}{\hbar}\int_{-\infty}^{t} E_n(t')dt'\right)$ with $j_c$ denoting the critical current density, $d^*$ the distance between two adjacent planes and $E_n$ the electric field between the n$^{th}$ and the (n+1)$^{th}$ superconducting planes. Due of this nonlinearity, and for a cavity tuned at $\nu_c > \nu_p$, we find by following [xxviii,xxix] that the resonance frequency of the cavity $\nu_c$ depends parametrically on the Josephson phase $\varphi_n$ ($\varphi_n = \chi_{n+1} - \chi_n$) as: $\nu_c(\varphi_n) = \sqrt{\nu_{c,0}^2 + \frac{d^*}{D}\nu_p^2\cos(\varphi_n)}$, where $\nu_{c,0}$ is the resonance frequency of the cavity without superconductor. Provided that a static current bias $j$ is applied along the c-axis, leading to a shift of the equilibrium phase difference $\varphi_n \to \varphi_n + \varphi_0$ (where $\sin(\varphi_0) = j/j_c$), the parametric coupling can be adjusted to vary between linear $\nu_c(\varphi_n) \propto \varphi_n$ and quadratic $\nu_c(\varphi_n) \propto \varphi_n^2$. The former case is formally equivalent to the interaction between the radiation pressure and the position of an end-mirror in a Fabry-Perot cavity [xxviii], typically used to cool macroscopic objects in optomechanics [viii]. Following the analogy with optomechanics, the quadratic coupling would describe the situation for a membrane-in-the-middle cavity, which has recently attracted much attention [xxx][xxxi]. In this situation, cooling and even squeezing have been predicted [xxxii].

Considering the lowest order only, the cavity resonance depends on the Josephson phase as $\nu_c(\varphi_n) \approx \nu_{c,0}(1 - \sin(\varphi_0)\frac{d^*}{2D}\frac{\nu_p^2}{\nu_{c,0}^2}\varphi_n)$ (where we have absorbed a small renormalization of the bare resonance frequency of the cavity $\nu_{c,0}$ originating from the static current bias $\nu_{c,0} \to \nu_{c,0}(1 + \cos(\varphi_0)\frac{d^*}{2D}\frac{\nu_p^2}{\nu_{c,0}^2})$)

The corresponding Hamiltonian reads :

$$H = h\nu_{c,0} a^+ a + h\nu_p b_n^+ b_n - hg(b_n^+ + b_n) a^+ a$$

Where $a^+, a, b_n^+$, and $b_n$ are the creation and annihilation operators of the photon and Josephson plasmon fields, respectively.

The coupling constant $g$ is expressed as: $g = \nu_{c,0} \sin(\varphi_0) \frac{d^*}{2D} \frac{\nu_p^2}{\nu_{c,0}^2} \varphi_n^{zpm}$ where $\varphi_n^{zpm} = \sqrt{\frac{h\nu_p}{2E_J\sqrt{\cos(\varphi_0)}}}$ is the zero point motion of the Josephson phase.

The above Hamiltonian describes the dynamics of a single Josephson phase $\varphi_n$ between a given set of neighboring planes that is coupled to the electromagnetic field of the cavity. In the situation considered here of a natural stack of Josephson junctions, we do have multiple superconducting planes and therefore a set of phases $\{\varphi_n\}_n$. Among the different normal modes of the stack, the in-phase mode, or "center of mass" mode, $B^+ = \frac{1}{\sqrt{N}} \sum_n b_n^+$ is the bright mode that couples to the cavity and can be cooled. Because direct coupling among the phases $\{\varphi_n\}_n$ originates from capacitive charge screening and is known to be weak compared to the coupling to the cavity mode rooted in the inductive screening of its magnetic field, we can consider them to be independent [xxxiii]. The center of mass mode retains its frequency $\nu_p$ and its dynamics is given by the same Hamiltonian as before :

$$H = h\nu_{c,0} a^+ a + h\nu_p B^+ B + h\sqrt{N} g(B^+ + B) a^+ a$$

but with the coupling constant renormalized as $g \to \sqrt{N} g$.

Let us estimate a threshold electric field at which cooling dynamics of the Josephson phase could be observed in this system. Cooling dynamics sets in when the effective damping rate of the Josephson phase $\Gamma_{J,eff}$ due to the back action of the cavity becomes comparable to the bare damping rate of the Josephson phase $\Gamma_J$ : $\frac{\Gamma_{J,eff}}{\Gamma_J} \sim 1$, where $\Gamma_{J,eff}$ is given by [xxxiv] : $\Gamma_{J,eff} = 4g^2 n/\Gamma_c$ with $n$ is the steady state number of photons in the cavity and $\Gamma_c$ the cavity damping rate. Introducing the Josephson quality factor $Q_J = \nu_p/\Gamma_J$, the quality factor of the cavity $Q_c = \frac{\nu_c}{\Gamma_c}$, the electromagnetic power enhancement inside the cavity $F = \frac{E_{cav}^2}{E_{in}^2}$ where $E_{in}$ and $E_{cav}$ are the strengths of electric field incident from vacuum and inside the cavity respectively, it amounts to :

$$\frac{\Gamma_{J,eff}}{\Gamma_J} = \frac{4g^2 n}{\Gamma_c \Gamma_J} = Q_J Q_c F \frac{\sin(\varphi_0)^2}{4\sqrt{\cos((\varphi_0))}} \frac{Nd^*}{D} \frac{\nu_p^2}{\nu_{c,0}^2} \left(\frac{2ed^* E_{in}}{h\nu_{c,0}}\right)^2$$

For a conservative choice of parameters such as $F = 10$, $Q_c = 10$, $Q_J = 10$ [xvii], $\frac{j}{j_c} = 0.95$, $\nu_{c,0} = 4$THz, $\nu_p = 1$THz, $Nd^* = D$ (filling fraction $f=1$), the phase cooling dynamics would set in at an incident electric field of $E_{in} \approx 20$kV/cm. This field strength is already available either with table top THz sources [xxxv] or at Terahertz-Free electron lasers. In addition, optimization of the geometry of these types of cavities could lead to further enhancement of the field inside the cavity with $F \approx 1000$ to $10000$ [xv] leading to a cooling dynamics accessible at much lower field thresholds. Our model describes the possibility of phase cooling below T$_c$. As for the choice of the superconductor, phase cooling could be achieved at any doping. Nevertheless, in the underdoped regime of the high-Tc superconductors, one could expect unconventional electronic properties resulting in macroscopic changes like the increase of T$_c$.

In summary, we have discussed how oxide heterostructuring can be used to generate arrays of optical cavities to dress and manipulate Josephson plasmons in high-Tc superconductors. The conditions discussed here may allow for experiments in which the interlayer phase fluctuations of high-Tc superconductors are cooled with light. More generally, we presented a versatile cavity design that is suitable for embedding the unconventional electronic order parameters of complex oxide materials. The combination of quantum materials with optical cavities provides a means to dress or even cool the eigenmodes of solids. This could open an exciting perspective of controlling the dynamics of order parameters at a phase transition, and may be the transition temperature itself.

## Acknowledgments

We are thankful to D. Jaksch,  S. Clark, S. Denny, A. Subedi, M. Eckstein, L. Zhang and R. Merlin  for discussions and suggestions, and J. Harms for support with the figures.  The research leading to these results has received funding from the European Research Council under the European Union's Seventh Framework Programme (FP7/2007-2013) / ERC Grant Agreement n° 319286 (Q-MAC).

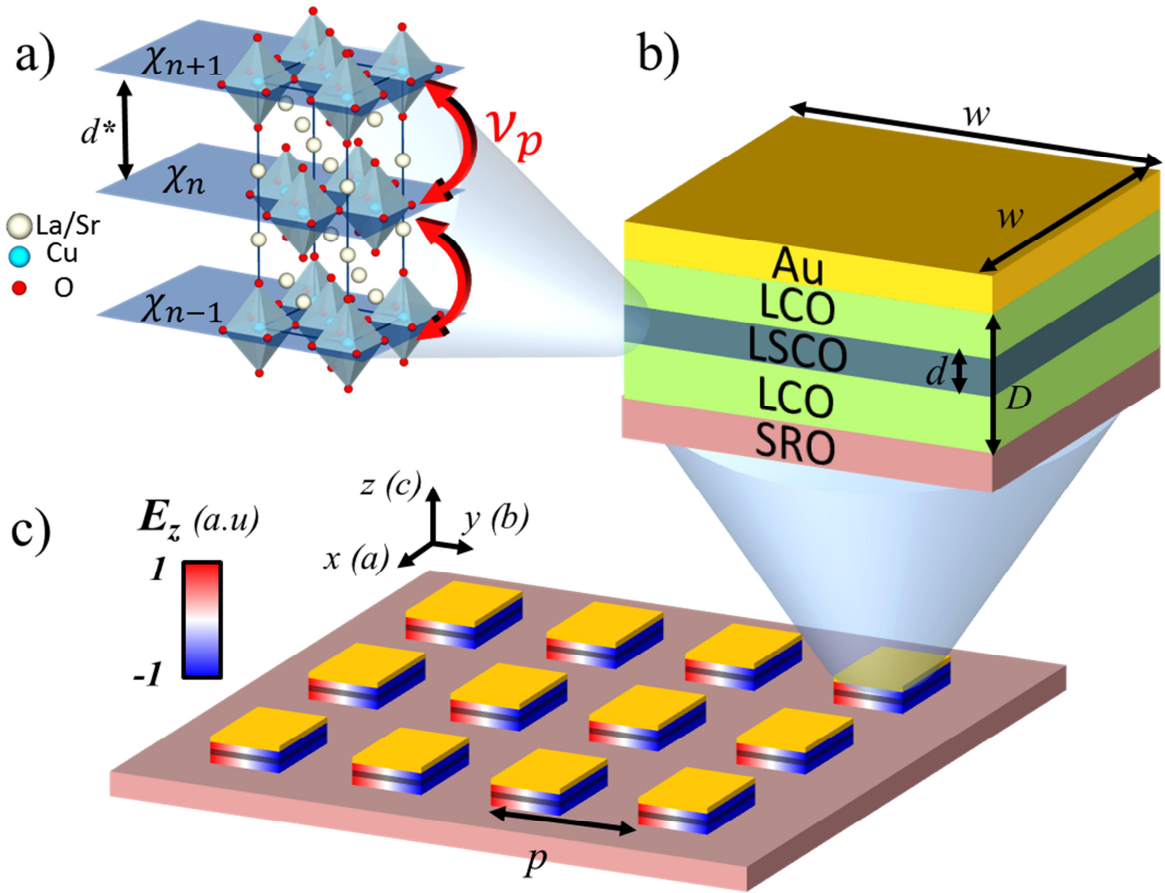

Figure 1 : **a)** the layered high-$T_c$ superconductor $La_{2-x}Sr_xCuO_4$: $\chi_n$ is the phase of the superconducting order parameter of the $n^{th}$ supeconducting plane. Red arrows indicate the Josephson plasma oscillations between the planes at the characteristic frequency $\nu_p$ **b)** a single cavity/high-$T_c$ mesa : Au (yellow)/$La_2CuO_4$("LCO",green)/$La_{2-x}Sr_xCuO_4$ ("LSCO", dark blue)/$La_2CuO_4$("LCO", green)/$SrRuO_3$("SRO", red) c) the proposed heterostructure : array of cavity/high-Tc mesas. The electric field distribution of the (n,m)=(0,1) fundamental mode of the "bare" cavities is shown (that is without the superconductor, see text for details). The corresponding notations are used thorough the text.

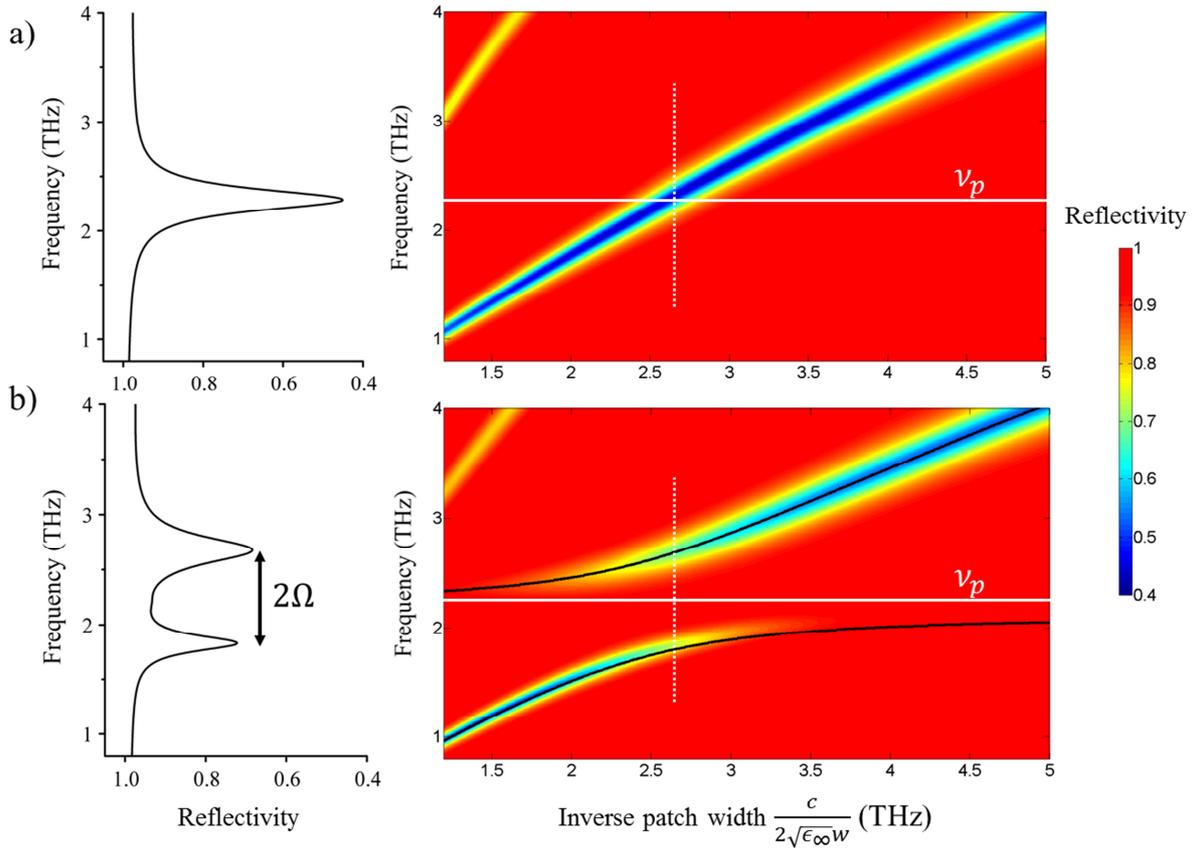

Figure 2 : **a)** <u>Right:</u> Reflectivity as a function of frequency and inverse patch width for the bare cavity (D=1.2µm, p=33 µm). The white solid line shows the JPR frequency $\nu_p$. <u>Left :</u> Reflectivity cut of the bare cavity tuned at $\nu_p$ (w=11 µm : corresponding to the white dashed line on the right panel) **b)** <u>Right:</u> Reflectivity as a function of frequency and inverse patch width for the cavity/high-$T_c$ heterostructure (d=200nm, D=1.2µm, p=33 µm, corresponding to a filling fraction $f = \frac{d}{D} \approx 17\%$). The white solid line shows the JPR frequency $\nu_p$. The black solid lines are the hybridized resonances $\{\nu_+(w), \nu_-(w)\}$ (see text for details) <u>Left:</u> Reflectivity cut of the cavity/ high-$T_c$ heterostructure when the bare cavity is tuned at $\nu_p$ (w=11 µm : corresponding to the white dashed line on the right panel)

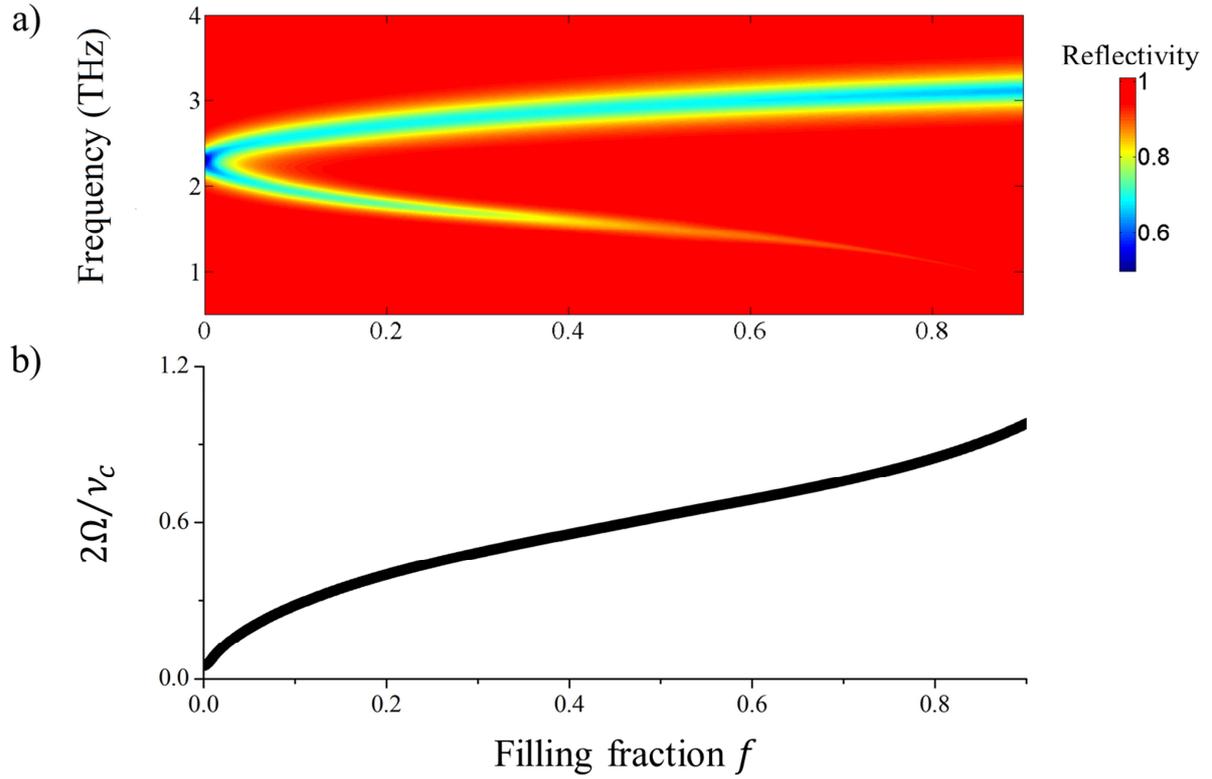

Figure 3 : **a**) Reflectivity as a function of frequency and filling fraction $f$ for a cavity tuned at $\nu_c = \nu_p$ (computed at fixed D=1.2μm, p=33 μm, w=11 μm) **b)** Normalized Rabi splitting $2\Omega/\nu_c$ as a function of the filling fraction $f$ computed from Figure 3a)